\documentclass[reqno]{article}
\usepackage{ae} 
\usepackage[T1]{fontenc}
\usepackage[ansinew]{inputenc}
\usepackage{amsmath}
\usepackage{amssymb}
\usepackage{graphicx}
\usepackage{color}
\usepackage{epsfig}

\newcommand{\beq}{\begin{equation}}
\newcommand{\eeq}{\end{equation}}
\newcommand\beqa{\begin{eqnarray}}
\newcommand\eeqa{\end{eqnarray}}
\newcommand\bea{\begin{array}}
\newcommand\eea{\end{array}}
\newcommand\ba{\begin{array}}
\newcommand\ea{\end{array}}
\newcommand{\nn}{\nonumber}

\newcommand{\neqa}{\nonumber\end{eqnarray}}
\newcommand{\la}[1]{\label{#1}}

\newcommand{\p}{\partial}
\newcommand{\pp}{{\rm p}}

\newcommand{\eq}[1]{eq.(\ref{#1})}
\newcommand{\rf}[1]{(\ref{#1})}
\newcommand{\eqs}[2]{eqs.(\ref{#1},\ref{#2})}

\newcommand{\hf}{\frac{1}{2}}

\renewcommand{\d}{\partial}
\renewcommand{\O}{{\cal O}}

\newcommand{\<}{{\langle}}
\renewcommand{\>}{{\rangle}}

\newcommand{\re}{\relax{\rm I\kern-.18em R}}

\renewcommand{\sp}{p\hspace{-.40em}/}

\def\su2{{SU(2)}}

\def\tr{{\rm tr}}

\def\s{{\rm s}}

\def\a{{\alpha}}

\def\({\left(}
\def\){\right)}
\def\[{\left[}
\def\]{\right]}

\def\e{\epsilon}

\def\s{\sigma}
\def\a{\alpha}
\def\b{\beta}
\def\th{\theta}

\def\({\left(}
\def\){\right)}
\def\[{\left[}
\def\]{\right]}

\def\<{\langle}
\def\>{\rangle}

\def\sG{\,\slash\!\!\!\! G}

\begin{document}

\renewcommand{\thefootnote}{\fnsymbol{footnote}}
\setcounter{footnote}{0}

\thispagestyle{empty}
\begin{flushright}
LPTENS-06/10\\
\end{flushright}
\vspace{.5cm}
\setcounter{footnote}{0}
\begin{center}
{\Large{\bf   Asymptotic Bethe Ansatz from String Sigma Model on $S^3\times R$} \vspace{4mm}  }\vspace{4mm}
{\large\rm Nikolay~Gromov$^{a,b}$, Vladimir~Kazakov$^{a,\!}$\footnote{Membre de
l'Institut Universitaire de France}\\[7mm]
\large\it\small ${}^a$ Laboratoire de Physique Th\'eorique\\
de l'Ecole Normale Sup\'erieure et l'Universit\'e Paris-VI,\\
Paris, 75231, France
\footnote{Unit\'e mixte de Recherche du
Centre National de la Recherche Scientifique et de  l'Ecole Normale
Sup\'erieure associ\'ee \`a l'Universit\'e de Paris-Sud et
l'Universit\'e  Paris-VI\\
\tt \indent\ \ gromov@thd.pnpi.spb.ru, nik\_gromov@mail.ru\\
 \indent\ \ kazakov@physique.ens.fr}\vspace{3mm}\\
\large\it\small ${}^b$ St.Petersburg INP,\\
 Gatchina, 188 300, St.Petersburg, Russia
}

\end{center}
\noindent\\[45mm]
\begin{center}
{\sc Abstract}\\[2mm]
\end{center}

We derive  the asymptotic Bethe ansatz (AFS equations
  \cite{Arutyunov:2004vx}) for the string on $S^3\times R$ sector of
  $AdS_5\times S^5$
  from the integrable nonhomogeneous dynamical spin
  chain for the string sigma model proposed in \cite{Gromov:2006dh}. It is clear
  from the derivation that AFS equations can be viewed only as an
  effective model describing a certain
  regime of a more fundamental inhomogeneous spin chain.

\newpage

\setcounter{page}{1}
\renewcommand{\thefootnote}{\arabic{footnote}}
\setcounter{footnote}{0}

\section{Introduction}

Recently,  it was proposed \cite{Gromov:2006dh} to describe the
compact, $R\times S^5$ sector of the Green-Schwarz-Metsaev-Tseytlin
(GSMT) superstring on $AdS_5\times S^5$ background by the
inhomogeneous, dynamical spin chain (we will abbreviate it to IDSC)
built out of the physical particles of the quantum $SO(6)$ sigma
model. The particles obey Zamolodchikovs factorizable S-matrix
\cite{Zamolodchikov:1977nu} and are put on the space circle
representing the worldsheet direction of the closed string in the
conformal gauge. Their isotopic $SO(6)$ vector degrees of freedom
realize the target space projection to $S^5$. The proposal of
\cite{Gromov:2006dh} was inspired by a similar construction of
\cite{Mann:2005ab}, who considered a different, supersymmetric
conformal $OSP(2m + 2|2m)$ sigma-model.

This sigma model is  asymptotically free and cannot pretend to the
precise quantum description of the GSMT superstring. However, as was
shown in \cite{Gromov:2006dh}, the Bethe ansatz equations of the
model perfectly reproduce in the classical limit the finite gap
solutions of \cite{KMMZ,Beisert:2004ag} in terms of their algebraic
curve\footnote{In \cite{Mann:2005ab} it was done for the
$OSP(2m+2|2m)$ sigma model, but the details of the classical limit
are there quite different from ours.}.

A natural question can be posed about the proposal of
\cite{Gromov:2006dh}: does it capture some essential  features of
the quantum superstring theory? For the moment we dispose a very
limited information about the quantum behavior of GSMT superstring.
The only robust calculations are made in the so called BMN limit
\cite{Berenstein:2002jq} where the quantum one loop
$1/\sqrt{\lambda}$\footnote{$\lambda$ is the 't Hooft coupling
constant of the dual N=4 SYM theory; $\hbar=1/\sqrt{\lambda}$ plays
the role of the Planck constant of the worldsheet  sigma model.}
corrections were calculated \cite{Callan:2003xr}, as well as the
same corrections around the classical solutions for the simplest
string motions: for the rotating circular and folded string
\cite{Frolov:2002av,Frolov:2003qc,Park:2005ji}.

In an interesting attempt to quantize the string on $R\times S^3$,
the authors of \cite{Arutyunov:2004vx} conjectured a discretization
of the classical finite gap equations of \cite{KMMZ} which in
addition has the right BMN and  the large gauge coupling limits. It
was later pointed out \cite{Schafer-Nameki:2005tn,Beisert:2005cw}
that the resulting equation for Bethe roots \eq{AFS} (the so called
AFS equation) can capture only a part of the truth, having chances
to describe only  large $\lambda$ and large angular momentum $L$ of
the string on $S^3$).

Nevertheless, the AFS equations  appeared to be a useful empirical
guideline for the search of quantum formalism based on
integrability. First, they appear to be the Bethe ansatz equations
of a certain non-local spin chain \cite{Beisert:2004jw}. Second,
they were generalized in a natural way to the full superstring
theory in \cite{Beisert:2005fw}, following nice observations of
\cite{Staudacher:2004tk}. This general model appears to be different
from the asymptotic SYM Bethe ansatz equation \eq{BDS}, the so
called BDS equation, proposed in  \cite{Beisert:2004hm}, the so
called BDS equation, only by a universal scalar factor $\sigma^2$
\eq{sigma2}. The BDS is better justified and to great extent even
deduced from the superalgebra of SYM theory
\cite{Beisert:2005wm,Beisert:2005tm}. The BDS equation is
claimed to be constrained by the crossing relations
\cite{Janik:2006dc}. It reproduces correctly at least three loops of
anomalous dimensions of SYM theory \cite{Staudacher:2004tk}.
Moreover, the quantum corrections of
\cite{Frolov:2002av,Frolov:2003qc,Park:2005ji} were reproduced on
the basis of AFS equations with some modifications in
\cite{Beisert:2005cw,Hernandez:2006tk,Arutyunov:2006iu,Freyhult:2006vr}.
All this means that the empirical  AFS equations  contain some grain
of truth about the structure of the quantum superstring.

The main problem with AFS equations is that they are not deduced
from any general enough principle as an approximation with respect
to a parameter, but rather proposed as an empirical guess.

In this paper, we will show that the AFS equation \eq{AFSBAE}
follows for large $\lambda$ and $L$ from the integrable quantum
inhomogeneous dynamical spin chain (IDSC) of \cite{Gromov:2006dh}
based on the sigma model on $S^3$.  The IDSC  plays similar role for
the AFS equation as the Hubbard model \cite{Rej:2005qt} for the BDS
equation of \cite{Beisert:2004hm}. The derivation is quite
straightforward and it demonstrates the qualitative nature of the
AFS equations, explicitly containing large parameters $L$ and
$\lambda$. The IDSC model seems to be more simple and fundamental
since it is a selfconsistent integrable quantum system. It may
incorporate the known data for both perturbative small $\lambda$,
SYM theory and the large $\lambda$ quasiclassical string results
since it has two \textit{a priory} adjustable functions, dispersion
of the particles and the scalar S-matrix factor, which can differ
from the Zamolodchikovs form and describe a different quantum
physics.

The paper is organized as follows. In section 2 we remind the
classical formulation of sigma model on $S^3$, its IDSC
quantization, as well as the AFS asymptotic string Bethe ansatz and
the asymptotic BDS SYM equations. In section 3 we derive the AFS
Bethe equations from the IDSC model and discuss a possible relation
to BDS equations. The section 4 is devoted to discussion and
conclusions. The details of the derivation of AFS formula are given
in the Appendices A,B. The Appendices C, D are devoted to the
generalization to the full scalar sector.\footnote{The result of
this generalization is different from the asymptotic Bethe equations
of \cite{Beisert:2005fw}, probably because it completely ignores the
interactions with fermions and the noncompact sector; however, the
formulas might be potentially useful for the search of the IDSC
formulation of the full GSMT superstgring. }
In Appendix E we give simplified derivation of the classical
limit of \cite{Gromov:2006dh}.

\section{  String on $S^3\times R$ space as inhomogeneous spin chain  }

\subsection{ The sigma-model on $S^3\times R$}

The action of the  $S^3\times R_t$ reduction of
Green-Schwarz-Metsaev-Tseytlin superstring  on $AdS_5\times S^5$
background \cite{Green:1981yb,Metsaev:1998it} in the conformal gauge
 in terms of homogeneous target space
coordinates $X_i(\tau,\s),\ i=1,\dots, 4$ and a scalar $Y(\tau,\s)$
representing the AdS time looks as follows\footnote{The coupling
constant in front of the action is identified by the AdS/CFT
correspondence with the 't Hooft coupling $\lambda=g^2N_c$ of the
${\cal N}=4$ supersymmetric Yang--Mills (SYM) theory
\cite{Frolov:2003qc}}:
\beq\label{SIGM}
S=\frac{\sqrt{\lambda}}{4\pi}\int_0^{2\pi}d\sigma\int d\tau\[(\d_a
X_i)^2-(\d_a Y)^2\],\;\;\;\;\;   X_iX_i=1. \eeq

The Virasoro conditions in the  gauge
\beq
 Y=\kappa\tau
\la{PERIODT}\eeq
read
\beq\label{VIRASORO}  (\d_\pm X_i)^2=(\p_\pm Y)^2=\kappa^2
\,\,\, , \,\,\, \d_\pm=\p_\tau\pm\p_\s\,, \eeq
$\Delta=\sqrt{\lambda}\,\kappa$ is identified with the dimension of
the corresponding SYM operator according to the AdS/CFT
correspondence.

After the gauge is imposed, the model looks like the  $O(4)$ sigma
model.  The action of the theory can be represented in terms of the
$SU(2)$ group valued field $\hat g=X_1+i\tau_3 X_2+i\tau_2 X_3+i\tau_1
X_4$ ($\tau_i$ are the Pauli matrices). Then the action \eq{SIGM}
takes the form of the $SU(2)$ principal chiral field
\beq\label{SIGMAM}
S=\frac{\sqrt{\lambda}}{4\pi}\int_0^{2\pi}d\sigma\int d\tau\(\d_a
X_i\)^2=-\frac{\sqrt{\lambda}}{8\pi}\int d^2\sigma\;\tr\(j^{L,R}_a
j^{L,R}_a\), \eeq
where
\beq  j^R_\pm= \hat g^{-1}\p_\pm \hat g\,,\qquad j^L_\pm= \d_\pm\hat g\hat g^{-1}\,,
\eeq
are the  currents of the  global symmetry of  left and right
multiplication by $SU(2)$ group element. The corresponding Noether
charges are
\beq Q_L=\frac{\sqrt{\lambda}}{4\pi}\int d\sigma\ \tr\( i\d_0 \hat g\,
\hat g^\dag \tau^3\),\;\;\;\;\; Q_R=\frac{\sqrt{\lambda}}{4\pi}\int
d\sigma\ \tr\( i \hat g^\dag \d_0 \hat g\, \tau^3\)\,.\;\;\;\;\; \eeq
%

 From the action we read off the energy and momentum as
\beq \la{EPclassic}
E\pm P=-\frac{\sqrt{\lambda}}{8\pi}\int\tr[j_0\pm
j_1]^2d\sigma=\frac{\sqrt{\lambda}}{2}\kappa^2\, \eeq
and in particular one has the level matching condition  \beq
\la{lmtch} P=0\,. \eeq

\subsection{ Inhomogeneous dynamical spin chain (IDSC)   }

Of course, the sigma-model \eq{SIGM} can be viewed as a consistent
truncation of the full GSMT superstring to the $S^3\times R$ sector
only classically. In the full quantum sigma-model one cannot turn
off the interactions with fermions, and through them, with all the
rest of the bosonic coordinates of the string.  On the other hand,
in the dual $N=4$ SYM theory the $SU(2)$ sector seems to be closed
at least perturbatively.   So one may hope that one could capture
the essential features of the quantum string sigma model by
considering  quantization of  the pure  $O(4)$ sigma model
\eq{SIGM}, or of the equivalent $SU(2)_L\times SU(2)_R$ principal
chiral field, \eq{SIGMAM}.

The asymptotically free  sigma-model \eq{SIGM} is integrable and
obeys the factorizable S-matrix for its physical particles
\cite{Zamolodchikov:1977nu} (see \cite{Gromov:2006dh} for the
details). It was proposed in \cite{Gromov:2006dh}, following the
analysis of a similar supersymmetric sigma-model in
\cite{Mann:2005ab}, to describe the quantum  string as a collection
of $L$ physical particles with rapidities $\theta_k,\quad
k=1,\cdots,L$, on a circle of the length $2\pi$ representing the
periodic world sheet $\sigma$-coordinate of the closed string.

The periodicity condition on the circle for the Bethe wave function
of these particles, written in terms of a system of Bethe equations
(see the details in \cite{Gromov:2006dh}),  looks as follows
\begin{eqnarray}
e^{-i\mu \sinh\pi\theta_\a}|\psi\rangle={\prod_{1}^{\a-1}}\ \hat
S\(\th_\a-\th_\b\) {\prod_{N}^{\a+1}}\ \hat
S\(\th_\a-\th_\b\) |\psi\rangle \label{MBAE}\,.
\end{eqnarray}
The r.h.s. of these equations has a form of a spin chain
transfer-matrix acting on an eigenvector $|\psi\rangle$ with
eigenvalue $e^{-i\mu \sinh\pi\theta_\a}$. Since the matrices $\hat
S$ describing the quantum spins at individual sites of such a spin
chain depend on $\theta$'s, which by themselves should be determined
from this equation, we call this system inhomogeneous dynamical spin
chain, or IDSC for short\footnote{We hope to avoid the confusion
with the dynamical spin chain with a changing length of
\cite{Beisert:2003ys}}.

The eigenvalue problem \eq{MBAE} can be solved by the standard
algebraic Bethe ansatz techniques  and gives the following system of
nested Bethe ansatz equations
\begin{eqnarray}
e^{-ip(\theta_\alpha) }&=& \prod_{\beta\neq \alpha}\, S_0^{\,2} \(
\th_\a-\th_\b \) \prod_j\frac{\th_\a-u_j+i/2}{\th_\a-u_j-i/2}\,
\prod_k\frac{\th_\a-v_k+i/2}{\th_\a-v_k-i/2}\,, \label{DBAE1} \\
1&=&\prod_\b\frac{u_j-\th_\b-i/2}{u_j-\th_\b+i/2}
\prod_{i\neq j} \frac{u_j-u_i+i}{u_j-u_i-i}\,,  \label{DBAE2}\\
1&=&\prod_\b\frac{v_k-\th_\b-i/2}{v_k-\th_\b+i/2} \prod_{l\neq k}
\frac{v_k-v_l+i}{v_k-v_l-i}\,, \label{DBAE3}
\end{eqnarray}
where
\begin{eqnarray*}
\a,\b=1,\dots,L,\;\;\;\;\;\;\;\;i,j=1,\dots,J_u,\;\;\;\;\;\;\;\;
k,l=1,\dots,J_v\,.
\end{eqnarray*}
For the pure $SO(4)$ sigma model  we should take the relativistic
momentum dispersion for the physical particles
\beq\label{DISPP} p(\theta_\alpha) =\mu \sinh\pi\theta_\a \eeq
and the  Zamolodchikovs S-matrix scalar factor
\begin{eqnarray}
S_0(\theta)=i\frac{\Gamma\(-\frac{\theta}{2i}\)\Gamma\(\frac{1}{2}
+\frac{\theta}{2i}\)}{\Gamma\(\frac{\theta}{2i}\)
\Gamma\(\frac{1}{2}-\frac{\theta}{2i}\)},\;\;\;\;\;
\la{S0S04}
\end{eqnarray}
where
 \beq \la{mug}\mu=e^{-\frac{\sqrt\lambda}{2}}=e^{-\frac{2\pi g}{\sqrt 2}}
\eeq
according to the asymptotic freedom.

Energy and momentum are given by  standard relativistic relations
\begin{eqnarray}
\la{P_SSL} P=\frac{\mu}{2\pi}\sum\sinh(\pi\theta_\alpha) \,,
\end{eqnarray}
\begin{eqnarray}
\la{E_SSL}E=\frac{\mu}{2\pi}\sum\cosh(\pi\theta_\alpha)\,.
\end{eqnarray}

In what follows we consider the limit of large $g$. Rapidities
$\theta_\alpha$ can be considered as  \textit{coordinates} of
particles in an external potential $\mu \cosh\pi\theta_\alpha$ with
interaction
\beq\label{ASSS}  i\log S^2_0(\theta)\simeq 1/\theta + \O(1/\theta^3)
\,.  \eeq
When $g$ is large $\mu$ is exponentially small according to \eq{mug}
and the external potential becomes a square box potential with
infinite walls at $\theta=\pm\sqrt{2}g$. Since the size of the box
is large we can leave only the first, "Coulomb" term in interaction
(\ref{ASSS}) between the particles. We will consider only the states
with all $\theta$'s having the same mode number $m$ (see
\eq{ABAE1}). It was argued in \cite{Gromov:2006dh} that such a
selection of the states corresponds to the absence of longitudinal
oscillations of the string, which is well seen at least in the
classical limit characterized by a big number of large $\theta$'s.
Let us note that the $\theta$ excitations correspond roughly to the
unphysical longitudinal  motions of the string (which are excluded
by the choice of the single mode number) and the magnon variables
correspond to the transverse motions, which is well seen at least in
the classical limit (see \cite{Gromov:2006dh}).

 Strictly speaking, we can trust the periodicity conditions
\eq{MBAE} only in the limit when the energies of individual
particles is large $E_\alpha=\frac{\mu}{2\pi} \cosh(\pi
\theta_\alpha) \to \infty$, up to the terms $\sim e^{-E_\a}$. In the
classical limit the total energy is large, but the energies of
individual particles  are in our case  small for large $\lambda$. In
that case the very notion of S-matrix looses its sense, since it is
valid only in the infinite space, or at least when the size of the
space is much bigger than the invers mass. Nevertheless the
classical limit of \cite{Gromov:2006dh} works  well\footnote{We
thank Al. and A.~Zamolodchikovs for the discussion on this point. }.
It might be that the physical interpretation in terms of the
worldsheet particles is not adequate for the string and the right
interpretation is in terms of the inhomogeneous dynamical quantum
spin chain with the transfer matrix given by the r.h.s. of
\eq{MBAE}, with the functions $p(\theta)$ and $S_0(\theta)$ yet to
fix.


\subsection{AFS and BDS equations and notations}

The AFS equations   for the energy  spectrum of the $S^3$ subsector of
the string on $AdS_5\times S^5$ \cite{Arutyunov:2004vx} and BDS
equations for the anomalous dimensions of $SU(2)$ subsector of
${\cal N}=4$ SYM theory \cite{Beisert:2004hm} look very similar.
 For YM it
reads
\beq \la{BDS}\(\frac{y_k^+}{y_k^-}\)^L= \prod_{j\neq k}^K
\frac{u_k-u_j+i}{u_k-u_j-i} , \eeq
for string one has to multiply r.h.s. by a ``dressing" factor
$\sigma^2$
\beq \la{AFS}\(\frac{y_k^+}{y_k^-}\)^L= \prod_{j\neq k}^K
\frac{u_k-u_j+i}{u_k-u_j-i} \sigma^2(u_j,u_k)\,, \eeq
where
\beq \la{sigma2}\sigma^2(u_j,u_k)=
\(\frac{1-1/(y_j^-y_k^+)}{1-1/(y_j^+ y_k^-)}\)^{-2} \(\frac{(y_j^-
y_k^--1)}{(y_j^-y_k^+-1)} \frac{(y_j^+ y_k^+-1)}{(y_j^+
y_k^--1)}\)^{2i(u_j-u_k)}\,. \eeq

$y^\pm_j$ are defined by means of  Zhukovsky transformation
 and its inverse
\beq\la{ZX}  Z(x)\equiv x+1/x, \qquad X(z)\equiv \hf\(
z+\sqrt{z^2-4}\),   \eeq
where by definition we take the branch of the square root, so that
$|X(z)|>1$ for $z\in \mathbb C$. Then \beq y^\pm_j\equiv
X\(\frac{u_j\pm i/2}{M}\), \eeq where $M$,  introduced in
\cite{Gromov:2006dh}, is \beq \la{eM}
M=\frac{g}{\sqrt{2}}=\frac{\sqrt{\lambda}}{4\pi}\,. \eeq
The systems of equations (\ref{BDS},\ref{AFS}) should be accompanied
by the periodicity (level-matching) condition
\beq
\la{p_afs0}\sum_j\log\frac{y^-_j}{y^+_j}=2\pi i m. \eeq
And finally,
\beq
\la{delta_afs0}\Delta=L+2iM\sum_{j=1}^K\(\frac{1}{y_j^+}-\frac{1}{y_j^-}\),
\eeq
giving us either the dimension of a SYM operator or the energy of a
quantum string state.

\section{  AFS equations from inhomogeneous spin chain }

If we put $J_v=0$ thus retaining  only the excitations corresponding
to the left charges in \eq{DBAE1}-(\ref{DBAE3}) we obtain a reduced
system of Bethe equations.
\begin{eqnarray}
e^{-ip(\theta_\alpha) }&=& \prod_{\beta\neq \alpha}\, S_0^{\,2} \(
\th_\a-\th_\b \) \prod_j\frac{\th_\a-u_j+i/2}{\th_\a-u_j-i/2}\,
\,, \label{RBAE1} \\
1&=&\prod_\b\frac{u_j-\th_\b-i/2}{u_j-\th_\b+i/2} \prod_{i\neq j}
\frac{u_j-u_i+i}{u_j-u_i-i}\,,  \label{RBAE2}
\end{eqnarray}
where
\begin{eqnarray*}
\a,\b=1,\dots,L,\;\;\;\;\;\;\;\;i,j=1,\dots,J\,.
\end{eqnarray*}
Its classical limit reproduces the KMMZ finite gap equations which
can be compared with the perturbatively closed $SU(2)$ sector of
SYM.

Now we will show that in the continuous limit for
$\theta$-distribution $L\to\infty$, and $g\sim L$, but for an {\it
arbitrary} number $J_u$ of magnon variables $u_k$ \footnote{We put
$J_v=0$, taking only left excitations in the pricipal chiral field
language. We take both types of excitations into account in Appendix
C. The full compact $SO(6)$ subsector  is  considered in Appendix
E.} the \eq{RBAE1}-(\ref{RBAE2}) together with \eqs{P_SSL}{E_SSL}
reproduce the  AFS asymptotic Bethe ansatz from
\cite{Arutyunov:2004vx}. It will be clear from the derivation that
the AFS equation plays only a role of an effective equation
interpolating between BMN limit and classical limit. One could, for
example, take into account $1/\sqrt{\lambda}$ corrections stemming
from the discreteness of the $\theta$'s which are missing in AFS
equation.

It will be important for us that in \eq{ASSS} we  take only the the
first term, corresponding to the two dimensional Coulomb repulsion,
and the $\theta$-variables will be confined in this limit,  due to
the dispersion \eq{DISPP}, in a square box with the vertical
infinite walls at $\theta=\pm \sqrt{2}g$. It is clear from this that
the characteristic $\theta$'s should in principle scale as
$\theta\sim\sqrt{\lambda}\to\infty$. The system
\eq{RBAE1}-(\ref{RBAE2}) becomes
\begin{eqnarray}
{\int\hspace{-0.38cm}-}\frac{\rho_\theta(z)dz}{w-z}+2\pi m&=&
-i\sum_j\log\frac{w M-u_j+i/2}{w M-u_j-i/2}\,,\;\;\;\;\;w\in(-2,2)\,, \label{ABAE1} \\
1&=&\prod_\b\frac{u_j-\th_\b-i/2}{u_j-\th_\b+i/2}
\prod_{k(\neq j)} \frac{u_j-u_k+i}{u_j-u_k-i} \label{ABAE2}\\
&&\a,\b=1,\cdots,L,\qquad k,j=1,\cdots, J.\nn
\end{eqnarray}
where $M=\frac{g}{\sqrt{2}}$. These equations should be accompanied
by a prescription that $\rho_\theta(z)$ has support $[-2,2]$ and has
$1/\sqrt{2\pm z}$ behaviour near edges of the support. This
prescription follows directly from \eq{DBAE1} in the
$L\rightarrow\infty$ limit. For more details see
\cite{Gromov:2006dh}.

The  big parameter $g$ enters  these equations thus inciting us to
expand w.r.t. $1/g$. But to reproduce the AFS equations we do not do
it. We could even assume  that  $g$ is finite, thus imposing the
rectangular shape of the potential with the finite length
$2\sqrt{2}g$. This is not possible with the massive relativistic
dispersion \eq{DISPP}, where only the large $g$  limit leads to the
rectangular shape of the potential, but it might be the adequate
choice of dispersion in the full string theory, which should have
the conformal symmetry in the conformal gauge. The similarity with
the BDS equation where $g$ is small, and not big, supports this
idea.

\subsection{Density}

Let us now calculate the density of distribution of $\theta$'s.
 Taking  $\log$ of the \eq{ABAE1} we
obtain:
\beqa   \sG_\theta(z)+2\pi
m=\sum_{j=1}^Ki\log\frac{M z-u_j-i/2}{M z-u_j+i/2},
\eeqa
Where
\beq  G_\th(z)=\frac{1}{M}\sum_{\a}\frac{1}{z-\theta_\a/M}=\int_{-2}^{2} \frac{ dz' \rho_\th(z')}{z-z'}
\eeq
and $\sG_\theta(z)$ is a real part of $G_\theta(z)$.
We can find $G_\theta(z)$ as a function of $u_j$.

Performing the inverse Zhukovsky map \eq{ZX} we obtain the equation
\beqa
\la{sGG}\sG_\th(z)+2\pi m=i\sum_{j=1}^K\(\log\frac{x-y_j^+}{x-y_j^-}+
\log\frac{x-1/y_j^+}{x-1/y_j^-}\)
\eeqa
where $y_j^\pm=X\(\frac{u_j\pm i/2}{M}\)$, $x=X(z)$. $X$ is defined
in \eq{ZX}.

Introducing
 \beq H(z)=G_\theta\(Z(x)\) \eeq
we obtain from \eq{sGG}
\beqa \frac{1}{2}\[H(x)+H(1/x)\]=-2\pi m+
i\sum_{j=1}^K\(\log\frac{x-y_j^+}{x-y_j^-}+\log\frac{x-1/y_j^+}{x-1/y_j^-}\)\,. \eeqa

 The solution of this equation,
with the right asymptotics at infinity
$H(1/\epsilon)=G_\theta\(1/\epsilon\)\simeq L/M \epsilon$, is as
follows
\beqa  H(x)=\frac{\frac{L}{2M}+2\pi
m}{x-1}+\frac{\frac{L}{2M}-2\pi
m}{x+1}+i\sum_{j=1}^{K}\[\frac{2x}{x^2-1}\(\frac{1}{y_j^+}-\frac{1}{y_j^-}\)
-\frac{2x^2\log \frac{y_j^+}{y_j^-}}{x^2-1}
+2\log\frac{y_j^+x-1}{y_j^-x-1}\] \label{HDEN}\eeqa

In \cite{Gromov:2006dh} it was shown that in classical limit
\eqs{P_SSL}{E_SSL} can be expressed through poles of $H(x)$ in
$x=\pm 1$. This result can be generalized for the case we consider
there (see Appendix A). Extracting the residues of $H(x)$ at the
poles $x=\pm 1$ we can see that
\beqa
\la{delta_afs}\Delta&=&L+2iM\sum_{j=1}^K\(\frac{1}{y_j^+}-\frac{1}{y_j^-}\)\\
\la{p_afs}P&=&\(m-\frac{i}{2\pi}\sum_{j=1}^K\log\frac{y_j^+}{y_j^-}
\)\Delta =0 \eeqa
\eq{delta_afs} is precisely the expression for the anomalous
dimension \eq{delta_afs0} and \eq{p_afs} gives precisely the  zero
momentum condition for the AFS equation.

We can also compute the density of $\theta$'s as the imaginary part
of the resolvent $G_\theta(z)$ \beq \la{rho_3}\rho_\theta(Z(x)) =
\frac{{\rm Im} G_\theta(Z(x))}{\pi}=\frac{i}{2\pi}\[H(x)-H(1/x)\]
\eeq

\subsection{ Derivation of AFS formula   }

In this section we will exclude $\theta$ variables from
(\ref{DBAE2},\ref{DBAE3}) using the $\theta$-density calculated
above, and obtain the AFS equation \eq{AFS}. We are trying here to
go the same way as the authors of \cite{Rej:2005qt}, where the
similar variables were excluded in favor of the magnon variables in
Lieb-Wu equations for the Hubbard model.

Let us now exclude $\theta$'s from  \eq{DBAE2}, using the result
\eq{HDEN}. Taking the $\log$ of \eq{DBAE2} we obtain
\beq
\sum_{j\neq k} \log\frac{u_k-u_j+i}{u_k-u_j-i}+2\pi i n_k=
\sum_\b \log\frac{u_k-\th_\b+i/2}{u_k-\th_\b-i/2}
\equiv i p_k
\eeq
rewriting $p_k$ through density we have \beq \la{defpk}i
p_k=M\int_{-2}^2 \log\frac{z-w_k^+}{z-w_k^-}\;\rho_\theta(z)dz \eeq
where $w_k^\pm=\frac{u_k\pm i/2}{M}$. The function $\rho_\theta(z)$
is given by \eqs{rho_3}{HDEN}. In Appendix A we perform the
integration and obtain the following result \beqa \nn i
p_k&=&\sum_j\[2\log\frac{y^-_k y^+_j(y^-_j y^+_k-1)}{y^+_k
y^-_j(y^+_j y^-_k-1)}
-2i(u_j-u_k)\log\frac{(y^-_j y^-_k-1)(y^+_j y^+_k-1)}{(y^-_j y^+_k-1)(y^+_j y^-_k-1)}\]\\
\la{pk}&-&2M\(\frac{1}{y^+_k}-\frac{1}{y^-_k}\)\[2\pi
m-i\sum_j\log\frac{y^+_j}{y^-_j}\]+L\log\frac{y^+_k}{y^-_k} \eeqa
It leads to the following equations
\beq
\la{AFSBAE}\(\frac{y_k^+}{y_k^-}\)^L=\prod_{j\neq k}^K\frac{y_k^+-y_j^-}{y_k^--y_j^+}\(\frac{1-1/(y_j^-
y_k^+)}{1-1/(y_j^+ y_k^-)}\)^{-1} \(\frac{(y_j^- y_k^--1)}{(y_j^-
y_k^+-1)}\frac{(y_j^+ y_k^+-1)}{(y_j^+ y_k^--1)}\)^{2i(u_j-u_k)}
\eeq
which precisely coincide with the AFS \cite{Arutyunov:2004vx}
\rf{AFS}, including the expressions for energy and momentum
\eqs{delta_afs}{p_afs}.

\subsection{Periodicity in the world sheet momentum }

Denoting ${\rm p_j}=-i\log\frac{y^+_j}{y^-_j}$, where
$y^{\pm}=X\(\frac{u(\pp)\pm i/2}{M}\)$, we rewrite \eq{sigma2}  in the
way originally proposed in \cite{Arutyunov:2004vx}
\beq
\sigma^2(u_j,u_k)=\exp\(2i\sum_{r=2}^\infty
M^{2r}(q_r(\pp_k)q_{r+1}(\pp_j)-q_r(\pp_j)q_{r+1}(\pp_k))\) \eeq The
functions $q_r(\pp)$ (charges) are given by
\beq
q_r(\pp)=M^{2-2r}\frac{2\sin[\frac{1}{2}(r-1)\pp]}{r-1}
\(\frac{\sqrt{1+16M^2\sin^2(\frac{1}{2}\pp)}}{4\sin\(\frac{1}{2}\pp\)}\)^{r-1}=
i\frac{M^{-r+1}}{r-1}\(y_+^{1-r}-y_-^{1-r}\) \eeq
We used here that  from the definition of momentum ${\rm p}$ it
follows
\beq u(\pp)=\frac{1}{2}\cot\frac{\pp}{2}\sqrt{1+16 M^2\sin^2\frac{\pp}{2}}
\eeq
The energy and momentum \eqs{delta_afs}{p_afs} look now as follows
\beqa
\Delta-L&=&\sum_{j=1}^K\(\sqrt{1+16 M^2\sin^2({\rm p}_j/2)}-1\)\\
P&=&\(m+\frac{\sum_j{\rm p}_j}{2\pi}\)\Delta=0\,, \eeqa
 Thus by integrating out $\theta$'s we obtain, instead of the
 relativistic dispersion relations
(\ref{P_SSL},\ref{E_SSL}),  the lattice dispersion relations in the
effective magnon momentum defined through the l.h.s of \eq{AFSBAE}
as a free phase of the  magnon: $e^{-i{\rm
p}_k}=\(\frac{y_k^+}{y_k^-}\)$.
All these formulas are periodic with respect to the magnon momentum
${\rm p}$, which was inspired in the AFS construction by the SYM
spin chain.

On the SYM side, this dispersion relation was obtained in \cite{Berenstein:2005jq} in large $g$ limit and was than reproduced on the basis of integrability
and supersymmetry in \cite{Beisert:2004hm,Rej:2005qt}.
On the string side, the large $\lambda$ limit of this dispersion relation was  reproduced in
 \cite{Hofman:2006xt} considering the  ''giant magnon" configuration of the string.


%
%
%
%

\section{ Discussion}

In this paper we derived the asymptotic string Bethe ansatz, AFS
equations, directly from a more fundamental model of inhomogeneous
dynamical spin chain (IDSC). The generalization to the full
superstring theory is left to be guessed, and it will probably
constrain further the possible properties of the IDSC model. The
unknown dispersion relation for the ``particles" constituting the
chain, as well the scalar factor $S_0$ might become some functions
of $\lambda$ and could have just a different behavior at strong and
week coupling. Hopefully it will correspond to the known
perturbative SYM and the quasiclassical string data.
 This is a
possible way to reconcile the apparent differences of two theories,
and in particular, to resolve the annoying  3-loop discrepancy. The
methods of the direct calculation the bare string S-matrix of
\cite{Klose:2006dd}, as well as the results of \cite{Roiban:2006yc}
might help a lot.

Let us also note that the periodicity in momentum  of elementary
``magnon" excitations  in the string theory discussed recently in
\cite{Hofman:2006xt} follow naturally from our construction since we
reproduced the AFS equation. This periodicity, revealing the
lattice structure of the IDSC model, stems in our approach from the
specific distribution of $\theta$-variables: they are confined in a
square box and the density has a characteristic inverse square root
behavior at the end points.

For example, in the absence of magnons the density of
$\theta$-variables is  given by
\beq
\la{U1dens}\rho_0(\theta)d\theta=\frac{L}{\pi}\frac{d\theta}{\sqrt{2g^2-\theta^2}},\qquad    \theta\in (-\sqrt{2}g,\sqrt{2}).
\eeq
In terms of the variable $\theta=\sqrt{2}g\sin q$ this density
simply becomes  constant on a circle, like in the similar occasion
in Hubbard model (see the Appendix of \cite{Rej:2005qt}):
\beq  \rho_q(q)dq =\frac{L}{\pi} dq,\qquad    q\in (0,2\pi). \eeq
The periodicity is a simple consequence of this behavior of the
density.

It is interesting to note that taking in \eq{RBAE1}-(\ref{RBAE2})
\beq p(\theta_\alpha) =L\( \arcsin \frac{\theta_\a}{\sqrt{2}
g}-\phi\) \eeq
and
\begin{eqnarray}
S_0(\theta)= -1
\end{eqnarray}
we obtain the Lieb-Wu equations for the Hubbard model with the
energy of a state given by
\beq    E= \frac{1}{g^2} \sum_{\a=1}^{L} \sqrt{2g^2-\theta_\a^2}
\eeq
 which reproduces, according to
\cite{Serban:2004jf}, the 3-loop (and may be all loop) anomalous
dimensions $\Delta=E-L$. It is particularly clear in terms of the
variable $q$.

The bound states, ``strings", of a few magnons observed in
\cite{Dorey:2006dq} can be also naturally incorporated into the IDSC
model \footnote{We thank P.Vieira for pointing it out to us.}.

An important check for the relevance of the IDSC model for the
description of the GSMT superstring could come from the calculations
of quantum $1/\sqrt{\lambda}$ corrections. The discreteness of
$\theta$ variables should give some specific contributions to the
first correction to the energy
\cite{Schafer-Nameki:2006gk,Schafer-Nameki:2005is}, together with the contributions from
the discreteness of magnons calculated in
\cite{Schafer-Nameki:2005tn,Beisert:2005cw,Arutyunov:2006iu,Hernandez:2006tk,Freyhult:2006vr}.

An important unresolved question is the generalization of our
construction to the full superstring theory. The ``particles" out of
which we make our inhomogeneous dynamical chain are yet to be
identified in the full theory. An important guess might come from
the system of asymptotic string AFS-type equations for the full
superstring theory written in \cite{Beisert:2005fw}.

\subsection*{Acknowledgements}

We would like to thank   I.~Kostov, K,~Sakai, D.~Serban,
A.~Tseytlin, A. and Al.~Zamolodchikovs, K.~Zarembo and especially
P.~Vieira for discussions. The work of V.K. was partially supported
by European Union under the RTN contracts MRTN-CT-2004-512194 and by
INTAS-03-51-5460 grant. The work of N.G. was partially supported by
French Government PhD fellowship and by RSGSS-1124.2003.2.

\section*{Appendix A, Formula for energy}

In this appendix we derive formula for energy \eq{delta_afs}
directly from \eq{E_SSL}. In fact it is a trivial generalization of
the  result of \cite{Gromov:2006dh}, but we include it for
completeness.

As it was shown in \cite{Gromov:2006dh} density of $\theta$'s
behaves as $1/\sqrt{2\pm z}$ for $z\sim \mp 2$. We define
$\kappa_\pm$ as follows \beq \la{asympt_p}\rho\simeq
\frac{2\kappa_{\pm}}{\sqrt{2\mp z}},\;\;\;\;\;z\rightarrow \pm 2
\eeq

We want to compute the sum
\beq \nn E\equiv \frac{\mu}{2\pi}\sum_\alpha
\cosh(\pi\theta_\alpha)\,, \eeq
but we  \textit{cannot} simply replace this sum by an integral and
use the asymptotic density for $\theta$'s to compute the energy.
This is because the main contribution for the energy comes from
large $\theta$'s, near the walls, where the expression for the
asymptotic density is no longer accurate.

We notice that the energy
is dominated by large $\th$'s where, with exponential precision, we
can replace $\cosh \pi\theta_\a$ by $\pm \sinh \pi \th_\alpha$ for
positive (negative) $\theta_\a$. Then,
\beq \la{ener} E= \sum_{z_\alpha>0} \frac{\mu}{\pi} \sinh\( \pi
z_\alpha M\)- \sum_{z_\alpha<0}\frac{\mu}{\pi}\sinh \(\pi
z_\alpha M\)\, , \nn \eeq
Where $z_\a=\theta_\alpha/M$.
Having a
sum of $\sinh \pi \theta_\a$ we can substitute each  of them by the
corresponding Bethe equation (\ref{DBAE1}) obtaining
\beq E\simeq \frac{i}{\pi}\!\!\!\!
\sum_{z_\b<0<z_\a}\!\!\!\!\log
S_0^2\(M\[z_\alpha-z_\beta\]\)
+\sum_{j,\alpha} \frac{i\,{\rm sign}(z_\alpha)}{\pi}\log\frac{z_\a-w_j^-}{z_\a-w_j^+}
+\sum_{\alpha}m\;{\rm sign}(z_\alpha)\nn\,. \eeq
Where $w_j^\pm=\frac{u_j\pm i/2}{M}$. Now we can
safely go to the  continuous limit since in the first term the
distances between $\xi$'s are now mostly of order the $1$.
This allows to rewrite the energy, with $1/M$ precision, as follows
\beqa
E&\simeq& -\frac{M}{\pi}\int_{-2}^a dz\int_{a}^2 dw \frac{\rho_\theta(z)\rho_\theta(w)}{z-w}
+\frac{iM}{\pi} \sum_j\int \rho_\theta(z)\log\frac{z-w_j^-}{z-w_j^+}{\rm sign}(z-a)dz \nn\\
&+&m M  \int \rho_\theta(w) {\rm
sign}(w-a)dw\label{startingpoint}
\eeqa
where $a=0$.  But now, due to \eq{ABAE1}, one can see that we can
take any $a\in (-2,2)$. Indeed, taking a derivative of r.h.s. of
\eq{startingpoint} with respect to $a$, using \eq{ABAE1}, we get
zero. Hence we can even send $a$ close to the wall: $a=-2+\epsilon$,
where $\epsilon$ is very small. Let us calculate the first term. The
main contribution to the integral comes from $-2\simeq z\sim w$ so
that we can use the asymptotics (\ref{asympt_p}) to get
\beq \nn-\frac{M}{\pi}\int_{-2}^{-2+\epsilon} dz\int_{-2+\epsilon}^2
dw \frac{\rho_\theta(z)\rho_\theta(w)}{z-w}\simeq
-\int_{-2}^{-2+\epsilon} dz\int_{-2+\epsilon}^2 dw \frac{4 M
\kappa_-^2}{\pi(z-w)\sqrt{2+z}\sqrt{2+w}}\simeq  2\pi M \kappa_-^2
\eeq
The remaining $3$ terms are very simple: since $a\simeq -2$ we can
simply drop the $sign$-functions inside the integrals and obtain
exactly the expression of the momentum in the continuous limit. We
arrive therefore at
\beqa E\simeq 2 M \kappa_-^2 \pi+\(mL-\sum_p n_p J_p \) \,.
\la{energy} \eeqa
where $K_p$ is a number of $u$'s with mode number $n_p$.
If we compute the $a$-independent integral (\ref{startingpoint}) near the
other wall, i.e. for $a=2-\epsilon$, we find
\begin{eqnarray*}
E\simeq 2 M \kappa_+^2 \pi-\(mL-\sum_p n_p J_p\) \,. 
\end{eqnarray*}
Therefore, equating the results one obtains the desired expressions
for the energy and momentum
\beq
E\pm P=2\pi\,M\,\kappa_\pm^2 \la{EP}
\eeq
through the data $\kappa_\pm$ at the singularities of the curve at
$z=\pm 2$. From \eq{EP} we see that $\kappa_+=\kappa_-=\kappa$ to ensure $P=0$.
We can also write
\beq
\Delta=\lambda^{1/4}\sqrt{2E}=4\pi M\kappa
\eeq
and we immediately get \eqs{delta_afs}{p_afs} from \eq{HDEN}.

\section*{Appendix B, Derivation of AFS formula for asymptotic string BAE's}
In this appendix we evaluate integral (\ref{defpk}) and obtain AFS BAE.

We can simplify expression for $H(x)$ (\ref{HDEN}) assuming that in \eq{p_afs} $P=0$
\beq
H(x)=-4\pi m+\frac{\Delta}{M}\frac{x}{x^2-1}+2i\sum_j\log\frac{y_j^+ x-1}{y_j^- x-1}
\eeq
we rewrite (\ref{defpk}) in $x$ variable
\beq
i p_k=-\frac{M}{2}\oint \frac{i}{2\pi}\(H(x)-H(1/x)\)\(\log\frac{x-y_k^+}{x-y_k^-}+\log\frac{x-1/y_k^+}{x-1/y_k^-}\)\(1-\frac{1}{x^2}\)dx
\eeq
where contour goes in counterclockwise direction
around unit circle, $y_k^\pm=X\(w_k^\pm\)$.
Note that terms with $H(1/x)$ are equal to that with
$H(x)$ after change of the variable $x\rightarrow 1/x$. So that
\beq
i p_k=M\oint H(x)\(\log\frac{x-y_k^+}{x-y_k^-}+\log\frac{x-1/y_k^+}{x-1/y_k^-}\)\(\frac{x^2-1}{x^2}\)\frac{dx}{2\pi i}
\eeq

Various terms are
\beqa
I_1&\equiv&\oint\(-4\pi m+\frac{\Delta}{M}\frac{x}{x^2-1}\)\log\frac{x-y_k^+}{x-y_k^-}\(\frac{x^2-1}{x^2}\)\frac{dx}{2\pi i}\\
I_2&\equiv&\oint\(-4\pi m+\frac{\Delta}{M}\frac{x}{x^2-1}\)\log\frac{x-1/y_k^+}{x-1/y_k^-}\(\frac{x^2-1}{x^2}\)\frac{dx}{2\pi i}\\
I_3&\equiv&2i\oint \log\frac{y_j^+ x-1}{y_j^- x-1}\log\frac{x-y_k^+}{x-y_k^-}\(\frac{x^2-1}{x^2}\)\frac{dx}{2\pi i}\\
I_4&\equiv&2i\oint \log\frac{y_j^+ x-1}{y_j^- x-1}\log\frac{x-1/y_k^+}{x-1/y_k^-}\(\frac{x^2-1}{x^2}\)\frac{dx}{2\pi i}
\eeqa
Integral $I_1$ can be calculated by residue in $x=0$, since $|y_k^\pm|>1$.
\beq
I_1=\frac{\Delta}{M}\log\frac{y_k^+}{y_k^-}-4\pi m\(\frac{1}{y^+_k}-\frac{1}{y^-_k}\)
\eeq
Similar $I_2$ and $I_4$ are given by residue at infinity.
\beqa
I_2&=&4\pi m\(\frac{1}{y^+_k}-\frac{1}{y^-_k}\)\\
I_4&=&-2 i\(\frac{1}{y^+_k}-\frac{1}{y^-_k}\)\log\frac{y^+_j}{y^-_j}
\eeqa
Calculation of $I_3$ is slightly more difficult. One can differentiate
it with respect to $y_j^+$ to kill one of the logarithms and
then calculate it by poles at $x=0$
\beq
\d_{y_j^+} I_3=2 i\log\frac{y^+_k}{y^-_k}+2 i\(\frac{1}{{y_j^+}^2}-1\)\log\frac{y^+_k y^+_j-1}{y^-_k y^+_j-1},\;\;\;\;\;
I_3=\int_{y_j^-}^{y_j^+}\d_{y_j^+} I_3 \;dy_j^+
\eeq
thus
\beq
\nn I_3=
2i\frac{u_j-u_k}{M}\log\frac{(y^+_j y^-_k-1)(y^-_j y^+_k-1)}{(y^+_j y^+_k-1)(y^-_j y^-_k-1)}
+\frac{2}{M}\log\frac{y^-_j y^+_k-1}{y^+_j y^-_k-1}
+2 i\((y_j^+-y_j^-)\log\frac{y^+_k}{y^-_k}-(y_k^+-y_k^-)\log\frac{y^+_j}{y^-_j}\)
\eeq
Finally
\beq
i p_k=M\sum_{a=1}^4 I_a=L \log\frac{y^+_k}{y^-_k}+\sum_j\(2\log\frac{1-1/y_j^- y_k^+}{1-1/y_j^+ y_k^-}
+2 i (u_j-u_k)\log\frac{(y^+_j y^-_k-1)(y^-_j y^+_k-1)}{(y^+_j y^+_k-1)(y^-_j y^-_k-1)}\)
\eeq
thus we prove \eq{pk} assuming $P=0$. This immediately leads to AFS BAE \eq{AFSBAE}.

\section*{Appendix C,  Full $S^3\times R$ sector}

We can easily integrate out $\theta$'s variables in the case of both nontrivial
chiralities $J_u,\;J_v\neq 0$. Resolvent of the $\theta$'s becomes
 \beqa\la{Huv} H(x)=\frac{\frac{L}{2M}+2\pi
m}{x-1}+\frac{\frac{L}{2M}-2\pi
m}{x+1}+i\sum_{j=1}^{K_u+K_v}\[\frac{2x}{x^2-1}\(\frac{1}{y_j^+}-\frac{1}{y_j^-}\)
-\frac{2x^2\log \frac{y_j^+}{y_j^-}}{x^2-1}
+2\log\frac{y_j^+x-1}{y_j^-x-1}\] \eeqa
where we denoted
\beqa
y_j^\pm&=&X\(\frac{u_j\pm i/2}{M}\),\;\;\;\;\;j=1,\dots, K_u\\
y_{l+K_u}^\pm\equiv \tilde y_{l}^\pm&=&X\(\frac{v_l\pm i/2}{M}\),\;\;\;\;\;l=1,\dots, K_v
\eeqa
Formulas for energy and momentum thus have the same form \eqs{delta_afs}{p_afs}.
$u$'s and $v$'s don't entangle.
For magnon momentum \eq{defpk} we obviously have
 \beqa i
p_k&=&\sum_{j=1}^{K_u+K_v}\[2\log\frac{y^-_k y^+_j(y^-_j
y^+_k-1)}{y^+_k y^-_j(y^+_j y^-_k-1)} -2i(u_j-u_k)\log\frac{(y^-_j
y^-_k-1)(y^+_j y^+_k-1)}{(y^-_j y^+_k-1)(y^+_j
y^-_k-1)}\]+L\log\frac{y^+_k}{y^-_k} \eeqa
and thus BAE takes the form
\beqa
\(\frac{y_k^+}{y_k^-}\)^L&=&\prod_{j=1}^{K_u}\frac{y_k^+-y_j^-}{y_k^--y_j^+}\(\frac{1-1/(y_j^-
y_k^+)}{1-1/(y_j^+ y_k^-)}\)^{-1}
\(\frac{(y_j^- y_k^--1)}{(y_j^- y_k^+-1)}\frac{(y_j^+ y_k^+-1)}{(y_j^+ y_k^--1)}\)^{2i(u_j-u_k)}\!\!\!\!\!\!\!\!\!\!\!\!\!\!\!\!\!\!\!\!\times\\
\nn&\times& \prod_{l=1}^{K_v}\(\frac{1-1/(\tilde y_l^-
y_k^+)}{1-1/(\tilde y_l^+ y_k^-)}\)^{-2} \(\frac{(\tilde y_l^-
y_k^--1)}{(\tilde y_l^- y_k^+-1)}\frac{(\tilde y_l^+
y_k^+-1)}{(\tilde y_l^+ y_k^--1)}\)^{2i(v_l-u_k)} \eeqa
and symmetrical equation for $v$'s.

\section*{Appendix D,  $S^5\times R$ sector}

We can also generalize then calculation of Appendix C on $SO(6)$ case.
Full BAE for this case are

\begin{eqnarray}
\nn
e^{-i\mu\sinh\frac{\pi\theta_\a}{2}}&=&\prod_{\beta\neq\alpha}^{L}S_0(\theta_\a-\theta_\b)
\prod_{j=1}^{K_2}\frac{\theta_\a-u^{(2)}_j+i/2}{\theta_\a-u^{(2)}_j-i/2} \\
\nn 1 &=& \prod_{j\neq
i}^{K_1}\frac{u^{(1)}_i-u^{(1)}_j+i}{u^{(1)}_i-u^{(1)}_j-i}
\prod_{j=1}^{K_2}\frac{u^{(1)}_i-u^{(2)}_j-i/2}{u^{(1)}_i-u^{(2)}_j+i/2}\\
\la{qBAE_SO6}\prod_{\alpha=1}^{L}\frac{u^{(2)}_i-\theta_\alpha+i/2}{u^{(2)}_i-\theta_\alpha-i/2}
&=& \prod_{j\neq
i}^{K_2}\frac{u^{(2)}_i-u^{(2)}_j+i}{u^{(2)}_i-u^{(2)}_j-i}
\prod_{j=1}^{K_3}\frac{u^{(2)}_i-u^{(3)}_j-i/2}{u^{(2)}_i-u^{(3)}_j+i/2}
\prod_{j=1}^{K_1}\frac{u^{(2)}_i-u^{(1)}_j-i/2}{u^{(2)}_i-u^{(1)}_j+i/2} \\
\nn 1 &=& \prod_{j\neq
i}^{K_3}\frac{u^{(3)}_i-u^{(3)}_j+i}{u^{(3)}_i-u^{(3)}_j-i}
\prod_{j=1}^{K_2}\frac{u^{(3)}_i-u^{(2)}_j-i/2}{u^{(3)}_i-u^{(2)}_j+i/2}
\end{eqnarray}
where \beqa S_0(\theta)
=-\frac{\Gamma\left(\frac{1}{4}-i\frac{\theta}{4}\right)\Gamma\left(\frac{1}{2}-i\frac{\theta}{4}\right)\Gamma\left(\frac{3}{4}+i\frac{\theta}{4}\right)\Gamma\left(1+i\frac{\theta}{4}\right)}
{\Gamma\left(\frac{1}{4}+i\frac{\theta}{4}\right)\Gamma\left(\frac{1}{2}+i\frac{\theta}{4}\right)\Gamma\left(\frac{3}{4}-i\frac{\theta}{4}\right)\Gamma\left(1-i\frac{\theta}{4}\right)}\,.
\eeqa
and for large $\lambda$ one should take $\mu=\e^{-\frac{\sqrt{\lambda}}{4}}$. From the first line in \eq{qBAE_SO6}
we see that denoting
\beq
i p_k=M\int_{-2}^2 \log\frac{z M-u_k^{(2)}-i/2}{z M-u_k^{(2)}+i/2}\;\rho_\theta(z)dz
\eeq
we again have \eq{pk} with $y_j^\pm=X\(\frac{u^{(2)}_j\pm i/2}{M}\)$ and thus
we have the following set of effective BAE equations for large $L\sim \sqrt\lambda$
\begin{eqnarray}
\nn 1 &=& \prod_{j\neq
i}^{K_1}\frac{u^{(1)}_i-u^{(1)}_j+i}{u^{(1)}_i-u^{(1)}_j-i}
\prod_{j=1}^{K_2}\frac{u^{(1)}_i-u^{(2)}_j-i/2}{u^{(1)}_i-u^{(2)}_j+i/2}\\
\(\frac{y_i^+}{y_i^-}\)^L
&=&
\prod_{j\neq i}^{K_2}\(\frac{u^{(2)}_i-u^{(2)}_j+i}{u^{(2)}_i-u^{(2)}_j-i}\,\sigma^2(u_j,u_i)\)
\prod_{j=1}^{K_3}\frac{u^{(2)}_i-u^{(3)}_j-i/2}{u^{(2)}_i-u^{(3)}_j+i/2}
\prod_{j=1}^{K_1}\frac{u^{(2)}_i-u^{(1)}_j-i/2}{u^{(2)}_i-u^{(1)}_j+i/2} \\
\nn 1 &=& \prod_{j\neq
i}^{K_3}\frac{u^{(3)}_i-u^{(3)}_j+i}{u^{(3)}_i-u^{(3)}_j-i}
\prod_{j=1}^{K_2}\frac{u^{(3)}_i-u^{(2)}_j-i/2}{u^{(3)}_i-u^{(2)}_j+i/2}
\end{eqnarray}
%

\section*{Appendix E, Classical limit}

Having at hands explicit formula for resolvent \eq{HDEN} we can shortly show how the
classical equations of \cite{KMMZ} appears from (\ref{DBAE1}-\ref{DBAE3}).
Classical limit corresponds to large number of excitations $J_u,J_v\sim L \sim M$
and finite mode numbers $n_i^u,\;n_i^v\sim 1$. In this limit Bethe roots
scale as $M$. We denote
\beq
w_i=\frac{u_i}{M},\;\;\;\;\;\tilde w_i=\frac{v_i}{M},\;\;\;\;\;y_i=X\(\frac{u_i}{M}\),\;\;\;\;\;\tilde y_i=X\(\frac{v_i}{M}\)
\eeq
The resolvent \eq{Huv} in this limit becomes
\beq
\la{Hcl}H(x)=\frac{\frac{L}{M}x+4\pi m}{x^2-1}-\frac{2}{x^2-1}\frac{1}{M}\sum_{j=1}^{K_u+K_v}\frac{1}{1/x-y_j}
\eeq
Expanding \eq{DBAE2}  we have
\beqa \pi n_k&=&-\frac{1}{2}H(y_k)
-\frac{1}{y^2_k-1}\frac{1}{M}\sum_{j\neq k}^{K_u}\frac{1}{1/y_k-y_j}
+\frac{y^2_k}{y^2_k-1}\frac{1}{M}\sum_{j\neq
k}^{K_u}\frac{1}{y_k-y_j} \eeqa
Using \eq{Hcl} and dropping $\O(1/M)$ terms we have
\beqa
\la{baecl}\pi n_k&=&\frac{\frac{L}{2M}y_k+2\pi m}{1-y_k^2}
+\frac{1}{y_k^2-1}\frac{1}{M}\sum_{j=1}^{K_v}\frac{1}{1/y_k-\tilde
y_j} +\frac{y^2_k}{y^2_k-1}\frac{1}{M}\sum_{j\neq
k}^{K_u}\frac{1}{y_k-y_j} \eeqa
 Defining \cite{Gromov:2006dh}
\beq
p(x)=\frac{\frac{L}{2M}x+2\pi m}{1-x^2}
+\frac{1}{x^2-1}\frac{1}{M}\sum_{j=1}^{K_v}\frac{1}{1/x-\tilde y_j}
+\frac{x^2}{x^2-1}\frac{1}{M}\sum_{j=1}^{K_u}\frac{1}{x-y_j}
\eeq
we can see that $p(x)$ satisfy classical equations of KMMZ \cite{KMMZ}, namely
\beqa
&&p(1+\epsilon)\simeq  p(-1+\epsilon)\simeq-\frac{\Delta}{4 M\epsilon}\\
&&p(\epsilon)=2\pi m+\frac{\epsilon}{2M}(L-2J_v)\\
&&p(1/\epsilon)=-\frac{\epsilon}{2M}(L-2J_u)\\
&&\sp(x)=\pi n_j,\;\;\;\;\;x\in {\cal C}
\eeqa
where $M=\sqrt{\lambda}/4\pi,\;\Delta=\sqrt\lambda\,\kappa$.


\end{document}